\documentclass{Interspeech}

\interspeechcameraready

\usepackage{mathtools}
\usepackage{todonotes}
\usepackage{subcaption}
\usepackage{tabularx}
\usepackage{array}

\newcommand{\DefinedAs}{\triangleq}
\newcommand{\ApproximatedAs}{\simeq}

\DeclareMathOperator*{\argmax}{arg\,max}

\newcommand{\Entropy}{\mathcal{H}}
\newcommand{\ConditionalEntropy}{\Entropy_A}
\newcommand{\ExpectationOver}[1]{\mathbb{E}_{#1}}

\newcommand{\Audio}{\mathbf X}

\newcommand{\Words}{\mathbf w}
\newcommand{\BestWords}{\Words^*}

\newcommand{\AdaptationData}{\mathcal A}

\newcommand{\NBestList}{\mathcal{W}_N}

\newcommand{\AdaptationDistribution}{p_A}

\newcommand{\Recogniser}{q}
\newcommand{\JointDistribution}{q_{A}}
\newcommand{\BestRecogniser}{\Recogniser^*}

\newcommand{\Normalisation}{Z}

\newcommand{\Zeros}{\mathbf 0}

\let\oldalign\align
\let\oldendalign\endalign

\renewenvironment{align}
  {\linenomathNonumbers\oldalign}
  {\oldendalign\endlinenomath}

\title{Robust Unsupervised Adaptation of a Speech Recogniser Using Entropy Minimisation and Speaker Codes}

\author[affiliation={}, equalcontribution]{Rogier C.}{van Dalen}
\author[affiliation={}, equalcontribution]{Shucong}{Zhang}
\author[affiliation={}, equalcontribution]{Titouan}{Parcollet}
\author[affiliation={}]{Sourav}{Bhattacharya}

\affiliation[nocounter]{AI Center Cambridge}{Samsung}{United Kingdom}
\email{\{r.vandalen,s1.zhang,t.parcollet,sourav.b1\}@samsung.com}


\begin{document}

\maketitle

\begin{abstract} 
Speech recognisers usually perform optimally only in a specific environment and need to be adapted to work well in another.
For adaptation to a new speaker, there is often too little data for fine-tuning to be robust, and that data is usually unlabelled.
This paper proposes a combination of approaches to make adaptation to a single minute of data robust.
First, instead of estimating the adaptation parameters with cross-entropy on a single error-prone hypothesis or ``pseudo-label'', this paper proposes a novel loss function, the conditional entropy over complete hypotheses.
Using multiple hypotheses makes adaptation more robust to errors in the initial recognition.
Second, a ``speaker code'' characterises a speaker in a vector short enough that it requires little data to estimate.
On a far-field noise-augmented version of Common Voice, the proposed scheme yields a 20\,\% relative improvement in word error rate on one minute of adaptation data, increasing on 10 minutes to 29\,\%.
\end{abstract}

\section{Introduction}

New environments, whether a new domain or a new speaker, often cause speech recogniser performance to degrade.
Adaptation can improve performance, but requires data from the appropriate environment.
This paper will focus on unsupervised adaptation to a new speaker, a scenario which brings two problems: first, the adaptation data is usually unlabelled, making adaptation sensitive to errors in the hypotheses; and second, the amount of data per speaker is often limited to a minute or so.

The standard method for dealing with the first problem, unsupervised data, is to run the speech recogniser unadapted to generate a single hypothesis which is used as ``pseudolabel''.
However, adaptation is not robust to errors in the pseudolabel.
Section \ref{section:minimum_entropy} will therefore propose to exploit the fact that a speech recogniser outputs a distribution over word sequences, by minimising the entropy of this distribution \cite{grandvalet-2004-semi-supervised, wang-2021-tent, zhang-2022-memo}.
Minimum-entropy adaptation has an effect similar to pseudolabel adaptation in the case where the recogniser is certain, but it will hedge its bets where there are multiple plausible hypotheses.
Minimum-entropy adaptation is therefore more robust.

The second problem, the small amount of adaptation data, brings the risk of overfitting the adaptation parameters.
A popular adaptation method for neural networks, low-rank adaptation (LoRA) \cite{hu-2022-lora}, may train only, say, 1\,\% of model parameters, but they still number in the hundreds of thousands.
Section~\ref{section:speaker_code} will propose to use speaker codes \cite{abdel-hamid-2013-fast,xue-2014-direct} to make adaptation more robust.
Speaker codes are small conditioning vectors, in this work of length 1024,  learned with backpropagation.
I-vectors and x-vectors \cite{tan-2016-learning, dehak-2011-front-end} are also small conditioning vectors, but computed with a model optimised for  e.g.\ speaker verification.
In contrast, speaker codes are learned with a speech recognition objective.

Section \ref{section:experiments} will show experimental results on a version of Common Voice, a dataset chosen because it has multiple utterances from each speaker.
To make it realistic for far-field recognition, background noise is added with room simulation.
In experiments, adapting the low-dimensional speaker codes is more robust than LoRA.
The best method of increasing robustness is to minimise the conditional entropy.

\subsection{Related work}

Outside of the speech literature, \cite{grandvalet-2004-semi-supervised} proposed minimising the conditional entropy for semi-supervised training, and \cite{wang-2021-tent, zhang-2022-memo} for adaptation.
This is what section~\ref{section:minimum_entropy} will apply to an encoder-decoder speech recogniser, where each output class is a word sequence.
For speech, there has been work that minimises the entropy not of the distribution over word sequences, but of intermediate distributions that crop up inside a CTC or Transducer speech recogniser: \cite{lin-2022-listen} minimises the entropy of per-frame label distribution of a CTC recogniser; \cite{kim-2023-sgem} minimises the entropy of distributions over the next token conditioned on partial hypotheses for CTC and Transducer recognisers.
There also exists work that finds multiple hypotheses for adaptation by perturbing the input \cite{teja-2021-test} or the model \cite{khurana-2021-unsupervised}.
Finally, there is work that is only superficially related and adds in training (not adaptation) a loss term involving the entropy over CTC frame alignments for a single hypothesis (not over hypotheses): \cite{variani-2023-alignment} improves performance by minimising and \cite{chang-2023-revisiting} by maximising this entropy.

Given the limited amount of data available for speaker adaptation, a long-standing question is how to select a small set of parameters to estimate robustly (see \cite{bell-2021-adaptation} for an overview).
In the neural-network era, fine-tuning has been proposed for subsets of model parameters \cite{huang-2021-rapid} or with linear transformations of inputs or of outputs of intermediate layers, with easily millions of parameters \cite{neto-1995-speaker-adaptation, li-2010-comparison} or with limited expressivity \cite{swietojanski-2016-learning}.
A recent proposal from general machine learning, low-rank adaptation (LoRA) \cite{hu-2022-lora}, gives flexibility yet a smaller number of parameters, though still often numbering in the hundreds of thousands.
In contrast, speaker embeddings, such i-vectors, x-vectors \cite{tan-2016-learning, dehak-2011-front-end} or bottleneck features \cite{huang-2015-investigation, hsu-2017-unsupervised}, have a smaller dimensionality, but are optimised e.g.\ for speaker verification instead of speech recognition, the task at hand.
Speaker codes \cite{abdel-hamid-2013-fast,xue-2014-direct} are computed with backpropagation to improve speech recognition, but are small like embeddings.
They were originally proposed to adapt the input for an otherwise frozen DNN-HMM system.
Section~\ref{section:speaker_code} retains only the core idea, which is to perform fine-tuning on a conditioning vector.

\section{Minimum-entropy adaptation}
\label{section:minimum_entropy}



One problem that makes adaptation of a speech recogniser less robust is the reliance of a single hypothesis per utterance.
Standard adaptation runs a speech recogniser on an utterance, treats the resulting hypothesis as a ``pseudolabel'', and then adapts the parameters to make the pseudolabel more likely.
The procedure is like adaptation of a HMM model, where this can be seen as a variational approximation.
However, current-day speech recognisers are discriminative models $\Recogniser(\Words | \Audio)$ that classify utterance $\Audio$ from adaptation data $\AdaptationData$ into word sequence $\Words$.
The effect of the same procedure is different: pseudolabel adaptation finds the model $\BestRecogniser$ that makes pseudolabel $\BestWords$ more likely:
\begin{subequations}
    \begin{align}
        \label{eq:pseudolabel:adaptation}
        &\BestRecogniser := \argmax_{\Recogniser} \sum_{\Audio \in \AdaptationData}
        \Recogniser(\BestWords(\Audio) \vert \Audio), && \text{(adaptation)}; \\
        \label{eq:pseudolabel:recognition}
        &\text{where~}
        \BestWords(\Audio) \DefinedAs \argmax_{\Words} \Recogniser(\Words \vert \Audio)
        && \text{(recognition)}
        .
    \end{align}%
    \label{eq:pseudolabel}%
\end{subequations}%
This procedure is sensitive to errors in the initial hypothesis, since these errors are likely to be reinforced.
This paper proposes to increase robustness by using multiple hypotheses, with as a loss function the conditional entropy \cite{grandvalet-2004-semi-supervised, wang-2021-tent, zhang-2022-memo}.

To express the conditional entropy mathematically, first produce an auxiliary joint distribution $\JointDistribution$ over audio and words by multiplying the distribution $\AdaptationDistribution$ of audio from the target domain and the speech recogniser:
\begin{align}
    \JointDistribution(\Audio, \Words)
        \DefinedAs \AdaptationDistribution(\Audio) \Recogniser(\Words \vert \Audio)
    .
\end{align}
The entropy of this distribution can be written as a sum of conditional entropy $\ConditionalEntropy(\Recogniser)$, to be minimised, and the entropy over the audio $\Entropy(\AdaptationDistribution)$, which is constant w.r.t.\ the parameters of $\Recogniser$:
\begin{align}
    &\Entropy(\JointDistribution) =
        - \int\sum_{\Words}
            \JointDistribution(\Audio, \Words) \log \JointDistribution(\Audio, \Words)
        \,\mathrm{d}\Audio
    \notag\\&
    =
        - \ExpectationOver{\Audio \sim \AdaptationDistribution} \bigg[ \sum_{\Words}
            \Recogniser(\Words \vert \Audio) \log (\AdaptationDistribution(\Audio) \Recogniser(\Words \vert \Audio))
        \bigg]
    \notag\\&
    =
        \underbrace{
        - \ExpectationOver{\Audio \sim \AdaptationDistribution} \bigg[ \sum_{\Words}
            \Recogniser(\Words \vert \Audio)
            \log \Recogniser(\Words \vert \Audio)
        \bigg]}_{\DefinedAs \ConditionalEntropy(\Recogniser) \text{, to be minimised}}
        +
        \Entropy(\AdaptationDistribution)
    \label{eq:conditional_entropy}
    .
\end{align}
The first term in \eqref{eq:conditional_entropy}, the negated expected value, needs to be approximated in two ways.
First, the expectation over the distribution $\AdaptationDistribution$ over audio from the target domain is approximated using the ``empirical distribution'', i.e.\ as a sum over the observed adaptation data $\AdaptationData$.
Second, the unfeasible sum over all possible word sequences $\Words$ will be restricted to run over just an $N$-best list $\NBestList$.
The N-best list can be found through beam search with the unadapted system.

Straightforwardly restricting the sum to an $N$-best list would open a loophole: the loss could be minimised by shifting all probability mass away from the $N$-best list.
To close this loophole, this paper proposes to re-normalise the sum.
Define the normalisation as a function of $\Recogniser$:
\begin{align}
    \Normalisation(\Audio) \DefinedAs
    \hspace*{-0.2em}
    \sum_{\Words \in \NBestList(\mathrlap{\Audio)}}
    \hspace*{-0.2em}
    \Recogniser(\Words \vert \Audio)
    .
\end{align}
The final approximate loss function with the loophole closed is
\begin{align}
    \ConditionalEntropy(\Recogniser) \ApproximatedAs
        - \frac1{\lvert \AdaptationData \rvert}
        \sum_{\Audio \in \AdaptationData}
        \frac1{\Normalisation(\Audio)}
        \hspace*{-0.15em}
        \sum_{\Words \in \NBestList\mathrlap{(\Audio)}}
        \hspace*{-0.15em}
            \Recogniser(\Words \vert \Audio)
            \log \Recogniser(\Words \vert \Audio)
    \label{eq:conditional_entropy:approximated}
    .
\end{align}
Now, the overall weight is always $1$, and  the $\log \Recogniser(\cdot)$ term penalises any reduction of mass on the $N$-best list.

\section{Adaptation using speaker codes}
\label{section:speaker_code}

\begin{figure}
    \begin{subfigure}{0.53\columnwidth}
        \centering
        \includegraphics{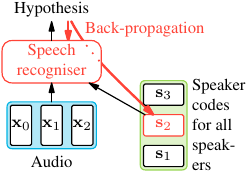}
        \caption{Jointly training the speech recogniser and all speaker codes.}
        \label{figure:speaker_code:training}
    \end{subfigure}
    \hfill
    \begin{subfigure}{0.43\columnwidth}
        \centering
        \includegraphics{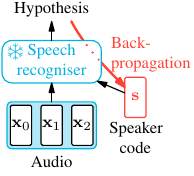}
        \caption{Adaptation to a speaker: fine-tuning the speaker code.}
        \label{figure:speaker_code:adaptation}
    \end{subfigure}
    \caption{Using speaker codes, at a high level.}
    \label{figure:speaker_code}
\end{figure}

The more parameters are estimated, the greater the risk of over-fitting.
Therefore, this paper adapts only 1024 parameters, contained in a vector called a ``speaker code'' \cite{abdel-hamid-2013-fast,xue-2014-direct}.
To make this handful of parameters meaningful, their space is optimised during training.
This has been called ``adaptive training'' \cite{gales-1998-cluster}.

Figure \ref{figure:speaker_code:training} shows how training with speaker codes works.
For each speaker in the training set, a separate speaker code vector is added to the trainable parameters.
The appropriate speaker code (for each speaker in the batch) is trained with backpropagation, jointly with the speech recogniser.
Thus, the speaker code space learns to capture inter-speaker variability, at the same time as the speech recogniser learns to exploit the speaker code to improve recognition.

In adaptation, in Figure \ref{figure:speaker_code:adaptation}, only the speaker code is fine-tuned, using a pseudolabel or the more sophisticated objective from section \ref{section:minimum_entropy}.
Before adaptation, the recogniser must come up with a hypothesis in a speaker-independent mode.
To support this, this paper uses multi-task learning to train the speech recogniser in speaker-independent and speaker-dependent mode at the same time.
For a fraction of utterances ($\frac12$ in the experiments), the speaker code is set to $\Zeros$ and not trained.

\begin{figure}
    \begin{subfigure}{0.48\columnwidth}
        \centering
        \includegraphics{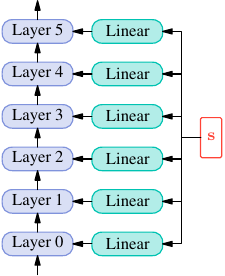} 
        \caption{Differently-transformed versions of the same speaker code are inserted into different Conformer layers, here layers 0 to 5.}
        \label{figure:injection:layers}
    \end{subfigure}
    \hfill
    \begin{subfigure}{0.48\columnwidth}
        \centering
        \includegraphics{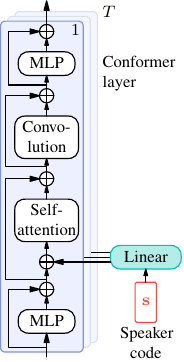} 
        \caption{Conformer layer, zoomed in.}
        \label{figure:injection:inside_residual}
    \end{subfigure}
    \caption{Injecting a speaker code into a Conformer speech recogniser.}
    \label{figure:architecture}
\end{figure}

The next question is how to inject the speaker code into the speech recogniser.
The experiments will use an encoder-decoder Conformer model.
Since this paper focuses on acoustic personalisation, the speaker code should affect the encoder, specifically the initial layers of the model.
From initial experiments with a few points of injection, the architecture illustrated in Figure \ref{figure:injection:inside_residual} worked best.
The speaker code is transformed with a separate linear projection (without bias) for each layer it is injected into (Figure \ref{figure:injection:layers}).
The linear layers are part of the model and not adapted.
In each of those Conformer layers (Figure \ref{figure:injection:inside_residual}), the transformed speaker code is added just before self-attention, after the residual connection branches off, enabling the model to bypass the speaker code if necessary.

\section{Experiments}
\label{section:experiments}

To evaluate speaker adaptation, a large data set that keeps track of speaker identities is necessary.
This work uses the Common Voice dataset \cite{ardila-2020-common_voice}, but augmented with noise and room simulation to make it similar to a realistic farfield scenario.
The recognition system is based on a SpeechBrain recipe \cite{ravanelli-2024-open-source}.

The Common Voice dataset, version 18, is used.
Speakers with less than 10 minutes of data are removed. The number of samples per speaker is limited to 5,000 as a few contributors recorded hundreds of hours of speech for Common Voice and we wish to limit biases towards them during ASR training. 

The ``train'' set has 1470 speakers, for 855 hours of data.
The ``train-dev'' and ``train-test'' sets, each 27 hours, are used to evaluate while training.
They contain the same speakers as the ``train'' set, so that not only the canonical model parameters but also the speaker codes can be evaluated during training.
The remaining 100 speakers are used for testing.
The ``adapt'', ``adapt-dev'' and ``test'' sets each contain 10 minutes of data from those speakers.
To confirm that there is no overlap, a SpeechBrain speaker recognition system%
    \footnote{\url{https://huggingface.co/speechbrain/spkrec-ecapa-voxceleb}}
is used to flag similar speakers, and it is manually confirmed that there is no overlap.
The ``adapt'' set will be used to perform adaptation, the ``adapt-dev'' set to determine the best epoch to use, and the ``test'' set for the word error rates.
Since the sets for adaptation and testing are separate, the amount of adaptation data from ``adapt'' can be varied while reporting word error rates on a consistent set: ``test''.

Background noises for augmentation are taken from the Musan dataset,%
\footnote{%
    David Snyder, Guoguo Chen, Daniel Povey,
    ``MUSAN: A Music, Speech, and Noise Corpus'', https://arxiv.org/abs/1510.08484
}
``music'' (with weight $\frac13$) and ``noise'' (with weight $\frac23$) subsets.
All music files are included, but noise files are selected based on length (at least 20 seconds), the variance in loudness across the waveform ($<20\,\text{dB}$), and the statistical skew of the loudness ($<-1$).
This leaves 88 files from the ``noise'' subset.
For each speaker, one type of background music/noise is randomly chosen, and added to all their data, repeating the noise where necessary, modelling a situation where any one speaker tends to use speech recognition in a particular environment, be it their car, an office, or on the street.
For one scenario, ``CommonVoice-Augmented'' the noise is added without reverberation.
For the main scenario, ``CommonVoice-Farfield'', the speech and background noise are rendered in a virtual room, roughly following \cite{kim-2017-generation}.
Rooms are sampled uniformly with horizontal sizes 1.5 to 20 metres, and height 1.5 to 5 metres.
The speech, the noise, and the virtual microphone are placed uniformly but at least 0.2 metre from any wall or floor.
The resulting distribution of ``RT60'', the time for the noise to drop by 60\,dB, is similar to the one in \cite{kim-2017-generation}.
After adding room reverb, levels are set to make the signal-to-noise ratio distributed according to $\mathrm{Unif}[\infty, 20, 10, 0]$.

The model is a 100M-parameter Conformer-based encoder-decoder system with joint CTC-Attention training \cite{kim-2017-joint}, implemented in SpeechBrain \cite{ravanelli-2024-open-source}.
The LoRA adaptation has with rank 16 into layers 1--5, for 1.3M parameters in total.
The speaker codes are injected, after initial experiments, as in Figure \ref{figure:injection:inside_residual}.
The models are trained for 175,000 steps with each batch containing 1,200 seconds based on dynamic duration batching.
For systems with speaker codes, a warm-up is performed for the first 5 epochs where speaker codes are kept at $\Zeros$.
After that, all parameters, including speaker codes and linear layers, are trained, using the Adam optimiser.
In each batch only speaker codes for the speakers in that batch are selected and trained.
The model thus jointly learns a meaningful space for speaker codes at the same time as the model that uses the speaker code.

However, one concern is the performance of the model before adaptation, i.e.\ with the speaker code set to $\Zeros$.
Therefore, multi-task training is used: for some of the utterances ($\frac12$ in this work), the speaker code is clamped to $\Zeros$, and not trained.

Adaptation and decoding is run independently per speaker.
For experiments with pseudolabels, the adaptation data ``adapt'' is decoded with the unadapted model.
Then, the adaptation parameters are fine-tuned, with all other parameters fixed and the reference transcription set to the pseudolabel.
For minimum-entropy training, the loss function is \eqref{eq:conditional_entropy:approximated}, where beam search yields a 5-best list for each utterance.
For all adaptation methods, the epoch out of 50 is found that maximises the unsupervised accuracy on the ``adapt-dev'', averaged over all speakers.
This number of epochs is then used to adapt the recogniser.
The word error rate on the ``test'' subset for each of the 100 speakers is then averaged.

\subsection{Results}

\begin{figure}
    \centering
    \includegraphics{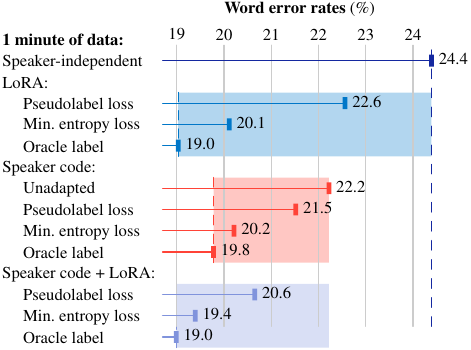}
    \caption{Average word error rates on CommonVoice-Farfield; adaptation on only one minute of data.}
    \label{figure:wers}
\end{figure}

The headline results for adaptation on only one minute of data are shown in Figure \ref{figure:wers}.
There are two aspects to the results: the adaptation parameters (LoRA, speaker code, or both) and the loss function (pseudolabel, minimum-entropy, or oracle label).
The shaded areas show the windows of opportunity for each choice of adaptation parameters, between unadapted and oracle label, i.e.\ supervised, adaptation.
Without adaptation, on CommonVoice-Farfield, a hard task, the word error rate is 24.4\,\%.
Intriguingly, training with speaker codes helps the system even if the speaker codes are set to zero (``Unadapted''), with performance at 22.2\,\%.
The hypothesis is that training with speaker codes has an effect like multi-task learning, but instead of having the speaker identity as an additional output, the speaker identity is given as a conditioning vector.

Two aspects stand out from Figure \ref{figure:wers}.
First, irrespective of which parameters are adapted, the minimum-entropy loss outperforms pseudolabel adaptation.
As explained in section~\ref{section:minimum_entropy}, the minimum-entropy loss exploits uncertainty in the recogniser output and thus makes adaptation robust to initial recognition errors.
When adapting both speaker codes and LoRA, the word error rate drops to 19.4\,\%, a 20\,\% relative improvement compared to the baseline.
Second, with the pseudolabel loss it is noticeable that adapting the 1024-dimensional speaker code, introduced in section \ref{section:speaker_code}, instead of LoRA is more robust.
The space of speaker codes, learned while training, is clearly a much terser description of speaker variation than LoRA matrices.

\begin{figure}
    \centering
    \includegraphics{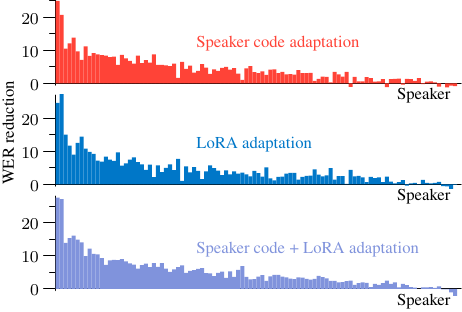}
    \caption{%
        Per-speaker improvement in word error rate from adapting with the minimum-entropy loss on one minute of data.
        Speakers are sorted by descending average improvement.
    }
    \label{figure:per_speaker}
\end{figure}

Figure \ref{figure:per_speaker} shows the improvement per speaker.
For all sets of adaptation parameters, performance improvements are seen for the vast majority of speakers, which underlines the robustness of the minimum-entropy loss.

\begin{figure}
    \centering
    \includegraphics{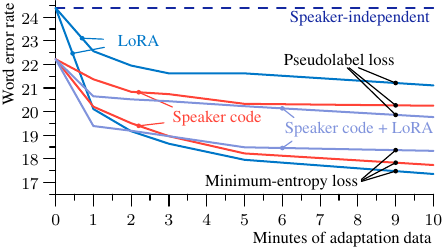}
    \caption{Average word error rates on CommonVoice-Farfield for adaptation with different amounts of adaptation data.}
    \label{figure:adaptation_data}
\end{figure}

Figure \ref{figure:adaptation_data} shows word error rates as a function of the amount of adaptation data.
The improvement that the minimum-entropy loss gives is consistent for all amounts of adaptation data.
For best performance, if there is a small amount of adaptation data, about one minute, it is useful to adapt both speaker code as well as LoRA.
As more adaptation data becomes available, it is worthwhile switching to just LoRA.

\subsection{Ablation studies}

\begin{figure}[t]
    \centering
    \includegraphics{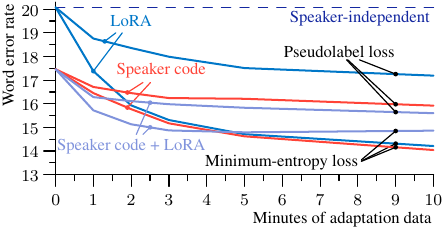}
    \caption{%
        Ablation: average word error rates on CommonVoice-Augmented for adaptation with different amounts of adaptation data.
        Unlike Figure~\ref{figure:adaptation_data}, the data has no reverberation.
    }
    \label{figure:adaptation_data:noise}
\end{figure}
\begin{figure}[t]
    \centering
    \includegraphics{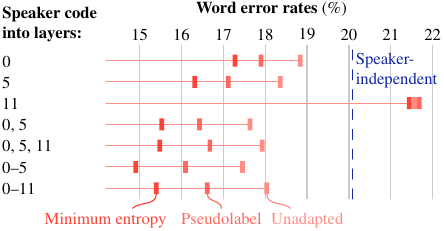}
    \caption{%
        Ablation: average word error rates on CommonVoice-Augmented for injecting the speaker code into different layers, for 5 minutes of adaptation data.
    }
    \label{figure:injection_layers}
\end{figure}

Figure \ref{figure:adaptation_data:noise}, like Figure \ref{figure:adaptation_data}, shows word error rates from different amounts of adaptation data, but on nearfield data from CommonVoice-Augmented.
In this scenario, minimum-entropy adaptation still provides significant improvements, but the harder scenario with reverberation, like in Figure \ref{figure:adaptation_data}, benefits more from the minimum-entropy loss.

Finally, Figure \ref{figure:injection_layers} examines recognition performance when the speaker code is injected into different blocks.
These results are on CommonVoice-Augmented again, and the procedure for choosing the adaptation epoch is different from the other experiments: here, for each speaker, the epoch is chosen to maximise the accuracy on that speaker's data from adapt-dev.
The expectation is that it is more important to inject information about the speaker in the lower layers, whereas in higher layers the differences between speakers are more likely to have been erased.
The results match this intuition: from Figure \ref{figure:injection_layers}, adapting the top layer, layer 11, consistently degrades performance, and in contrast the lower layers 0--5 are important for effective adaptation.

\section{Conclusion}
\label{section:conclusion}

This paper has examined how to adapt a speech recogniser to an unseen speaker, using only a small amount of unlabelled adaptation data.
It has proposed two methods to improve robustness of adaptation.
First, instead of using the standard pseudolabel loss, a new loss, the conditional entropy over the sequence of words that the speech recogniser outputs has been proposed.
Second, it has proposed to use speaker codes on an encoder-decoder recogniser.
Both methods help improve performance on small amounts of data.

\bibliographystyle{IEEEtran}
\bibliography{bibliography}

\begin{thebibliography}{10}
\providecommand{\url}[1]{#1}
\csname url@samestyle\endcsname
\providecommand{\newblock}{\relax}
\providecommand{\bibinfo}[2]{#2}
\providecommand{\BIBentrySTDinterwordspacing}{\spaceskip=0pt\relax}
\providecommand{\BIBentryALTinterwordstretchfactor}{4}
\providecommand{\BIBentryALTinterwordspacing}{\spaceskip=\fontdimen2\font plus
\BIBentryALTinterwordstretchfactor\fontdimen3\font minus
  \fontdimen4\font\relax}
\providecommand{\BIBforeignlanguage}[2]{{%
\expandafter\ifx\csname l@#1\endcsname\relax
\typeout{** WARNING: IEEEtran.bst: No hyphenation pattern has been}%
\typeout{** loaded for the language `#1'. Using the pattern for}%
\typeout{** the default language instead.}%
\else
\language=\csname l@#1\endcsname
\fi
#2}}
\providecommand{\BIBdecl}{\relax}
\BIBdecl

\bibitem{grandvalet-2004-semi-supervised}
Y.~Grandvalet and Y.~Bengio, ``Semi-supervised learning by entropy
  minimization,'' in \emph{Advances in Neural Information Processing Systems},
  2004.

\bibitem{wang-2021-tent}
D.~Wang, E.~Shelhamer, S.~Liu, B.~Olshausen, and T.~Darrell, ``Tent: Fully
  test-time adaptation by entropy minimization,'' in \emph{Proceedings of
  Conference on Learning Representations}, 2021.

\bibitem{zhang-2022-memo}
M.~Zhang, S.~Levine, and C.~Finn, ``{MEMO}: Test time robustness via adaptation
  and augmentation,'' in \emph{Advances in Neural Information Processing
  Systems}, 2022.

\bibitem{hu-2022-lora}
E.~J. Hu, Y.~Shen, P.~Wallis, Z.~Allen-Zhu, Y.~Li, S.~Wang, L.~Wang, and
  W.~Chen, ``Lo{RA}: Low-rank adaptation of large language models,'' in
  \emph{Proceedings of Conference on Learning Representations}, 2022.

\bibitem{abdel-hamid-2013-fast}
O.~Abdel-Hamid and H.~Jiang, ``Fast speaker adaptation of hybrid {NN/HMM} model
  for speech recognition based on discriminative learning of speaker code,'' in
  \emph{Proceedings of International Conference on Acoustics, Speech, and
  Signal Processing}, 2013.

\bibitem{xue-2014-direct}
S.~Xue, O.~Abdel-Hamid, H.~Jiang, and L.~Dai, ``Direct adaptation of hybrid
  {DNN/HMM} model for fast speaker adaptation in {LVCSR} based on speaker
  code,'' in \emph{Proceedings of International Conference on Acoustics,
  Speech, and Signal Processing}, 2014.

\bibitem{tan-2016-learning}
S.~Tan and K.~C. Sim, ``Learning utterance-level normalisation using
  variational autoencoders for robust automatic speech recognition,'' in
  \emph{Proceedings of Spoken Language Technology Workshop}, 2016.

\bibitem{dehak-2011-front-end}
N.~Dehak, P.~J. Kenny, R.~Dehak, P.~Dumouchel, and P.~Ouellet, ``Front-end
  factor analysis for speaker verification,'' \emph{IEEE/ACM Transactions on
  Audio, Speech, and Language Processing}, vol.~19, no.~4, 2011.

\bibitem{lin-2022-listen}
G.-T. Lin, S.-W. Li, and H.~Lee, ``Listen, adapt, better {WER}: Source-free
  single-utterance test-time adaptation for automatic speech recognition,'' in
  \emph{Proceedings of Interspeech}, 2022.

\bibitem{kim-2023-sgem}
C.~Kim, J.~Park, H.~Shim, and E.~Yang, ``{SGEM}: Test-time adaptation for
  automatic speech recognition via sequential-level generalized entropy
  minimization,'' in \emph{Proceedings of Interspeech}, 2023.

\bibitem{teja-2021-test}
P.~{Teja Sivaprasad} and F.~Fleuret, ``Test time adaptation through
  perturbation robustness,'' in \emph{NeurIPS Workshop on Distribution Shifts},
  2021.

\bibitem{khurana-2021-unsupervised}
S.~Khurana, N.~Moritz, T.~Hori, and J.~{Le Roux}, ``Unsupervised domain
  adaptation for speech recognition via uncertainty driven self-training,'' in
  \emph{Proceedings of International Conference on Acoustics, Speech, and
  Signal Processing}, 2021.

\bibitem{variani-2023-alignment}
E.~Variani, K.~Wu, D.~Rybach, C.~Allauzen, and M.~Riley, ``Alignment entropy
  regularization,'' in \emph{Proceedings of International Conference on
  Acoustics, Speech, and Signal Processing}, 2023.

\bibitem{chang-2023-revisiting}
O.~Chang, D.~Hwang, and O.~Siohan, ``Revisiting the entropy semiring for neural
  speech recognition,'' in \emph{Proceedings of Conference on Learning
  Representations}, 2023.

\bibitem{bell-2021-adaptation}
P.~Bell, J.~Fainberg, O.~Klejch, J.~Li, S.~Renals, and P.~Swietojanski,
  ``Adaptation algorithms for neural network-based speech recognition: An
  overview,'' \emph{IEEE Open Journal of Signal Processing}, vol.~2, 2021.

\bibitem{huang-2021-rapid}
Y.~Huang, G.~Ye, J.~Li, and Y.~Gong, ``Rapid speaker adaptation for {C}onformer
  {T}ransducer: Attention and bias are all you need,'' in \emph{Proceedings of
  Interspeech}, 2021.

\bibitem{neto-1995-speaker-adaptation}
J.~Neto, L.~Almeida, M.~Hochberg, C.~Martins, L.~Nunes, S.~Renals, and
  T.~Robinson, ``Speaker-adaptation for hybrid {HMM-ANN} continuous speech
  recognition system,'' in \emph{Proceedings of Eurospeech}, 1995.

\bibitem{li-2010-comparison}
B.~Li and K.~C. Sim, ``Comparison of discriminative input and output
  transformations for speaker adaptation in the hybrid {NN/HMM} systems,'' in
  \emph{Proceedings of Interspeech}, 2010.

\bibitem{swietojanski-2016-learning}
P.~Swietojanski, J.~Li, and S.~Renals, ``Learning hidden unit contributions for
  unsupervised acoustic model adaptation,'' \emph{IEEE/ACM Transactions on
  Audio, Speech, and Language Processing}, vol.~24, no.~8, 2016.

\bibitem{huang-2015-investigation}
H.~Huang and K.~C. Sim, ``An investigation of augmenting speaker
  representations to improve speaker normalisation for dnn-based speech
  recognition,'' in \emph{Proceedings of International Conference on Acoustics,
  Speech, and Signal Processing}, 2015.

\bibitem{hsu-2017-unsupervised}
W.-N. Hsu, Y.~Zhang, and J.~Glass, ``Unsupervised domain adaptation for robust
  speech recognition via variational autoencoder-based data augmentation,'' in
  \emph{Proceedings of Automatic Speech Recognition and Understanding
  Workshop}, 2017.

\bibitem{gales-1998-cluster}
M.~J.~F. Gales, ``Cluster adaptive training for speech recognition,'' in
  \emph{Proceedings of the International Conference on Spoken Language
  Processing}, 1998.

\bibitem{ardila-2020-common_voice}
\BIBentryALTinterwordspacing
R.~Ardila, M.~Branson, K.~Davis, M.~Kohler, J.~Meyer, M.~Henretty, R.~Morais,
  L.~Saunders, F.~Tyers, and G.~Weber, ``Common {V}oice: A
  massively-multilingual speech corpus,'' in \emph{Proceedings of the Language
  Resources and Evaluation Conference}, May 2020. [Online]. Available:
  \url{https://aclanthology.org/2020.lrec-1.520/}
\BIBentrySTDinterwordspacing

\bibitem{ravanelli-2024-open-source}
\BIBentryALTinterwordspacing
M.~Ravanelli, T.~Parcollet, A.~Moumen, S.~de~Langen, C.~Subakan, P.~Plantinga,
  Y.~Wang, P.~Mousavi, L.~D. Libera, A.~Ploujnikov, F.~Paissan, D.~Borra,
  S.~Zaiem, Z.~Zhao, S.~Zhang, G.~Karakasidis, S.-L. Yeh, P.~Champion,
  A.~Rouhe, R.~Braun, F.~Mai, J.~Zuluaga-Gomez, S.~M. Mousavi, A.~Nautsch,
  H.~Nguyen, X.~Liu, S.~Sagar, J.~Duret, S.~Mdhaffar, G.~Laperri{\`e}re,
  M.~Rouvier, R.~D. Mori, and Y.~Est{\`e}ve, ``Open-source conversational {AI}
  with {S}peech{B}rain 1.0,'' \emph{Journal of Machine Learning Research},
  vol.~25, no. 333, pp. 1--11, 2024. [Online]. Available:
  \url{http://jmlr.org/papers/v25/24-0991.html}
\BIBentrySTDinterwordspacing

\bibitem{kim-2017-generation}
C.~Kim, A.~Misra, K.~Chin, T.~Hughes, A.~Narayanan, T.~N. Sainath, and
  M.~Bacchiani, ``Generation of large-scale simulated utterances in virtual
  rooms to train deep-neural networks for far-field speech recognition in
  {G}oogle {H}ome,'' in \emph{Proceedings of Interspeech}, 2017.

\bibitem{kim-2017-joint}
S.~Kim, T.~Hori, and S.~Watanabe, ``Joint {CTC}-attention based end-to-end
  speech recognition using multi-task learning,'' in \emph{2017 IEEE
  international conference on acoustics, speech and signal processing
  (ICASSP)}.\hskip 1em plus 0.5em minus 0.4em\relax IEEE, 2017, pp. 4835--4839.

\end{thebibliography}

\end{document}